\documentclass[aip,10pt,floatfix,twocolumn,showpacs]{revtex4-1} 
\usepackage{color,soul}
\usepackage{graphicx}
\usepackage{siunitx}
\usepackage[breaklinks=true,colorlinks=true,linkcolor=blue,urlcolor=blue,citecolor=blue]{hyperref}

\usepackage{amsmath, amsthm, amssymb}
\usepackage{amsthm}
\usepackage{multirow}
\usepackage[normalem]{ulem}
\usepackage{soul}
\usepackage{makecell}
\usepackage[table]{xcolor}
\definecolor{Gray}{gray}{0.85}
\definecolor{LightCyan}{rgb}{0.88, 1, 1}
\definecolor{Apricot}{rgb}{0.98, 0.81, 0.69}
\usepackage{threeparttable}
\newcommand{\be}{\begin{equation}}
\newcommand{\ee}{\end{equation}}
\newcommand{\bea}{\begin{eqnarray}}
\newcommand{\eea}{\end{eqnarray}}

\usepackage[rightcaption,]{sidecap}
\sidecaptionvpos{figure}{c}

\begin{document}

\title{
Onset of glassiness in two-dimensional ring polymers: interplay of stiffness and crowding
}
\author{Sayantan Ghosh}
\email{sayantang@imsc.res.in}
\affiliation{The Institute of Mathematical Sciences, C.I.T. Campus,
Taramani, Chennai 600113, India}
\affiliation{Homi Bhabha National Institute, Training School Complex, Anushakti Nagar, Mumbai, 400094, India}
\author{Satyavani Vemparala}
\email{vani@imsc.res.in}
\affiliation{The Institute of Mathematical Sciences, C.I.T. Campus,
Taramani, Chennai 600113, India}
\affiliation{Homi Bhabha National Institute, Training School Complex, Anushakti Nagar, Mumbai, 400094, India}
\author{Pinaki Chaudhuri}
\email{pinakic@imsc.res.in}
\affiliation{The Institute of Mathematical Sciences, C.I.T. Campus,
Taramani, Chennai 600113, India}
\affiliation{Homi Bhabha National Institute, Training School Complex, Anushakti Nagar, Mumbai, 400094, India}
\date{\today}
\begin{abstract}
The effect of ring stiffness and pressure on the glassy dynamics of a thermal assembly of two-dimensional ring polymers is investigated using extensive coarse-grained molecular dynamics simulations. In all cases, dynamical slowing down is observed with increasing pressure and thereby a phase space for equilibrium dynamics is identified in the plane of obtained monomer density and ring stiffness. When the rings are highly flexible, i.e. low ring stiffness, glassiness sets in via crowding of crumpled polymers which take a globular form. In contrast, at large ring stiffness, when the rings tend to have large asphericity under compaction, we observe the emergence of local domains having orientational ordering, at high pressures. Thus, our simulations highlight how varying the deformability of rings leads to contrasting mechanisms in driving the system towards the glassy regime. 

\end{abstract}
\maketitle

\section{Introduction}\label{SEC-1}
Structural glass transition in most materials is  achieved either by lowering the temperature or by increasing the packing fraction of the constituent particles~\cite{binder2011glassy}.  In the latter context, focus has been mainly on hard sphere models and attempts at obtaining glassy behavior via differences in particle sizes, compositional ratios etc~\cite{parisi2010mean}. However these hardsphere models cannot capture the complex behavior of deformable particles of which colloid systems are a prime examples. Several works have focused on understanding the role of deformability in achieving disordered amorphous structures as a function of the packing fraction of such deformable objects in both two and three dimensions ~\cite{mukhopadhyay2011packings,makse2000packing,batista2010crystallization,boromand2018jamming,gnan2019microscopic}. It is possible that the introduction of deformability and softness of the particles can induce phases at high packing density that are not achievable with hard sphere particles. 

Ring polymers, as well as other topologically constrained polymers, have long been regarded as model systems for understanding the compaction and organization of biologically relevant polymers such as chromatin. One unique constraint imposed on closed loop polymers is that the primary mode of relaxation available for linear polymers, reptation, is unavailable for ring polymers ~\cite{ge2016self,halverson2014melt}. This alteration in dynamics and kinetics of high-density systems leads to certain arrested states in parameter space ~\cite{michieletto2016topologically, michieletto2017glassiness}, which is not the case for linear and unconstrained polymer systems. 

Polymer systems, with the potential for soft and tunable interactions, have been extensively used as a means to mimic soft colloids in numerous published works. These include linear polymers~\cite{louis2000can}, star or dendritic polymers~\cite{gotze2004tunable}, and more recently, topologically constrained ring polymers~\cite{gnan2019microscopic}. Polymer systems are uniquely placed as model systems to investigate various aspects of soft materials due to the existence of tunable interactions at multiple levels. They allow for the exploration of stability, rheology, and dynamics of different emergent phases.

The interactions between such soft and deformable polymers can vary, ranging from a coarse-grained description of interactions between the centers of mass to local bending interactions along the polymer backbone. These interactions dictate the stiffness of the polymer, which plays a significant role in the overall packing conformations and, ultimately, the material properties of such soft systems. Typically, jamming and glassiness in soft materials have been investigated using overlap-based soft sphere models \cite{o2003jamming, van2009jamming, chaudhuri2010jamming, ikeda2012unified}, which do not involve shape changes of the particles. However, in models where particles can undergo shape changes, both overlap and deformation can occur depending on various parameters, which provides exciting possibilities for packings and phase behaviour~\cite{manning2023essay}. In Voronoi-based models, the shape change is entirely determined by the particle centers. In vertex models, the shapes are constrained by dimensional factors. Both Voronoi-based models and vertex models are only applicable to confluent systems. However, by modeling particles as deformable polygons or ring polymers, it becomes possible to investigate both confluent and non-confluent systems, and the particle shape can independently influence the physics~\cite{boromand2018jamming}. It has been demonstrated~\cite{Loewe2020} that the mechanisms of overlap and deformation compete with each other, thereby influencing phase behaviour. 

The dimensionality also plays a critical role in the glassy behaviour observed in thermal ring polymers, as certain aspects such as threading and looping, which introduce additional time scales to the dynamics of the system, are not possible in 2D~\cite{vsiber2013many,miller2011two}. It is important, however, to compare the properties of similar systems in both 2D and 3D and take note of the differences. Previous work~\cite{gnan2019microscopic} has demonstrated that in 2D, soft colloidal particles modeled as ring polymers exhibit a novel 're-entrant melting', at very high packing fractions, driven by deformability of the rings. In this regime, there is a transition from fragile to strong behavior as the stiffness decreases. Earlier simulations of 2D disks have also shown that, under compression, the interaction between the constituent particles changes. This results in a transition from local deformation to a more distributed deformation field across the system at high compressions~\cite{vsiber2013many}. Furthermore, more recently, in the athermal limit, the jamming and mechanical response of two-dimensional systems of deformable polygons and the consequent shape changes have been analyzed~\cite{boromand2018jamming, treado2021bridging}. These findings now need to be connected to thermal observations in the context of properties around the glassy regime. 

In this work, we investigate the onset of slow dynamics leading to glassiness  in a 2D system of deformable polymer rings with increased compaction, for a range of ring stiffness. By monitoring the dynamical and structural aspects of these systems, we demonstrate how emergence of structural ordering spanning larger length scales in thermal assemblies of stiff ring polymers leads to arrested states in  dense deformable systems, in contrast to highly flexible polymers where crowding is the source of observed dynamical slowing down. The manuscript is organised as follow. After the introductory discussion in section I, the details of the model studied, the simulation as well as the measurements are provided in Section II. Our findings related to the dynamical and structural properties of the thermal assembly of two dimensional ring polymers is discussed in Section III. Finally, we provide a concluding discussion in Section IV.

\section{Modeling and Methods}\label{SEC-II}

\subsection{Model}

For the interaction between monomers, we used a combination of Weeks-Chandler-Andersen (WCA) potential
\begin{equation}
    U_{LJ}(r)=
    \begin{cases}
    4\epsilon[(\frac{\sigma_{m}}{r})^{12}-(\frac{\sigma_{m}}{r})^{6}]+\epsilon & \text{if }r\le 2^{\frac{1}{6}}\sigma_{m}\\
    0 & \text{if }r>2^{\frac{1}{6}}\sigma_{m}
    \end{cases}
\end{equation}
between all monomers, and finite extensible nonlinear elastic (FENE) potential
\begin{equation}
    U_{FENE}(r)=-\epsilon k_{F}R_{F}^{2}\ln [1-(\frac{r}{R_{F}\sigma_{m}})^{2}] \quad \text{if }r<R_{F}\sigma_{m}
\end{equation}
between bonded monomers, where $\sigma_{m}$ is the diameter of each monomer (and the unit of length), $\epsilon$ is the unit of energy, $k_{F}=15$ is the spring constant, and $R_{F}=1.5$ is the maximum extension of the bond ~\cite{smrek2020active}. To model the flexibility of the ring, we have used an angular potential
\begin{equation}
    V(\theta)=K_{\theta}(1-cos(\theta-\pi))
\end{equation}
In our study, we investigate how the variation of angular stiffness $K_{\theta}$ changes the dynamical and structural properties of the assembly of ring polymers. 

\subsection{Methods}

For our study, we consider a system of $N=100$ polymer rings, each consisting of $n_{m}=100$ monomers. To initialize the system, we randomly position the non-overlapping polymers. We then perform equilibration using NPT ensemble at a temperature of $T=1.0$ and the target pressure. The Nose-Hoover barostat is employed to maintain the desired pressure. After the system reaches equilibrium at the target pressure, we switch to the NVT ensemble to study the equilibrium dynamics and structure at the average density corresponding to the targeted pressure. Before commencing the production runs, we allow the system to run for $5 \times 10^{6}$ steps. Subsequently, the production run continues for approximately $10^{9}$ steps. The timestep is set to 0.01, and it is important to note that the pressure fluctuates around the mean value during these production runs. The molecular dynamics (MD) simulations were conducted using LAMMPS \cite{plimpton1995fast}. All quantities reported in this study are expressed in reduced LJ (Lennard-Jones) units.

In our study, we investigate multiple state points by varying the bending stiffness parameter, $K_{\theta}$, in the range of 1 to 20. These state points cover a wide range of pressures, leading to significant variations in density, as discussed in detail later in the paper. It is worth mentioning that a recent study conducted by Gnan et al. \cite{gnan2021dynamical} explored the properties of a similar model, referred to as SFPR. However, their study focused on smaller rings with $n_m=10$ (with some size polydispersity) and a limited variation in $K_\theta$ (ranging from 4 to 7).

\subsection{Observables}

\subsubsection{Dynamics}

To study the dynamics of the system, we mainly look at the mean-squared displacement (MSD) of the centers of mass (c.o.m) of the rings, which is given by $\Delta r^{2}(t)=\frac{1}{N}\sum_{i=1}^{N}|\Vec{r_{i}}(t)-\Vec{r_{i}}(0)|^{2}$ where $N$ is the total number of rings and $\Vec{r_{i}}(t)$ and $\Vec{r_{i}}(0)$ are the co-ordinates of the c.o.m. of the $i^{th}$ ring at times 0 and $t$ respectively. We measure the ensemble averaged MSD, $\langle\Delta r^{2}(t)\rangle$, from which the diffusion coefficient $D=\lim_{t\to\infty}\frac{\langle\Delta r^{2}(t)\rangle}{4t}$ is extracted.


\subsubsection{Structure}
The shapes of the rings within the assembly undergo changes as the density and stiffness vary. These changes can be quantified by measuring the asphericity of the rings and observing how it evolves with density. To compute the asphericity, we utilize the gyration tensor, denoted as $S_{mn}$. It is calculated using the formula: $S_{mn} = \frac{1}{N}\sum_{i=1}^{N}(r_{m}^{(i)}-r_{m}^{(CM)})(r_{n}^{(i)}-r_{n}^{(CM)})$, where $m$ and $n$ represent Cartesian coordinate indices. In two dimensions, each ring corresponds to a gyration tensor with two eigenvalues, $\lambda_{1}^2$ and $\lambda_{2}^2$. The asphericity of a ring, denoted as $a$, can then be defined as follows:

\begin{equation}
a= \frac{(\lambda_{1}^{2}-\lambda_{2}^{2})^2}{(\lambda_{1}^{2}+\lambda_{2}^{2})^2}
\end{equation}

We also measure the radius of gyration of the rings, $R_{g}$, which can be obtained from the eigenvalues:

\begin{equation}
    R_{g}=\sqrt{\lambda_{1}^{2}+\lambda_{2}^{2}}
\end{equation}

To quantify the presence of orientational order among the rings in our system, we calculate the orientational correlation function, as described by Frenkel et al. \cite{Frenkel2000}. The orientational correlation function, denoted as $g_l(r)$, is defined as the average of the cosine of the angle difference between the orientations of two rings separated by a distance $r$. In our study, we specifically focus on the second-order orientational correlation function, $g_2(r)$, which measures the correlations related to possible nematic order among the rings. By analyzing $g_2(r)$, we can gain insights into the extent of orientational correlations and the presence of nematic ordering within the system.

\section{Results}\label{SEC-III}
\subsection{Effect of stiffness on dynamical properties}

\begin{figure}[t]
\centerline{\includegraphics[width=1.0\columnwidth]{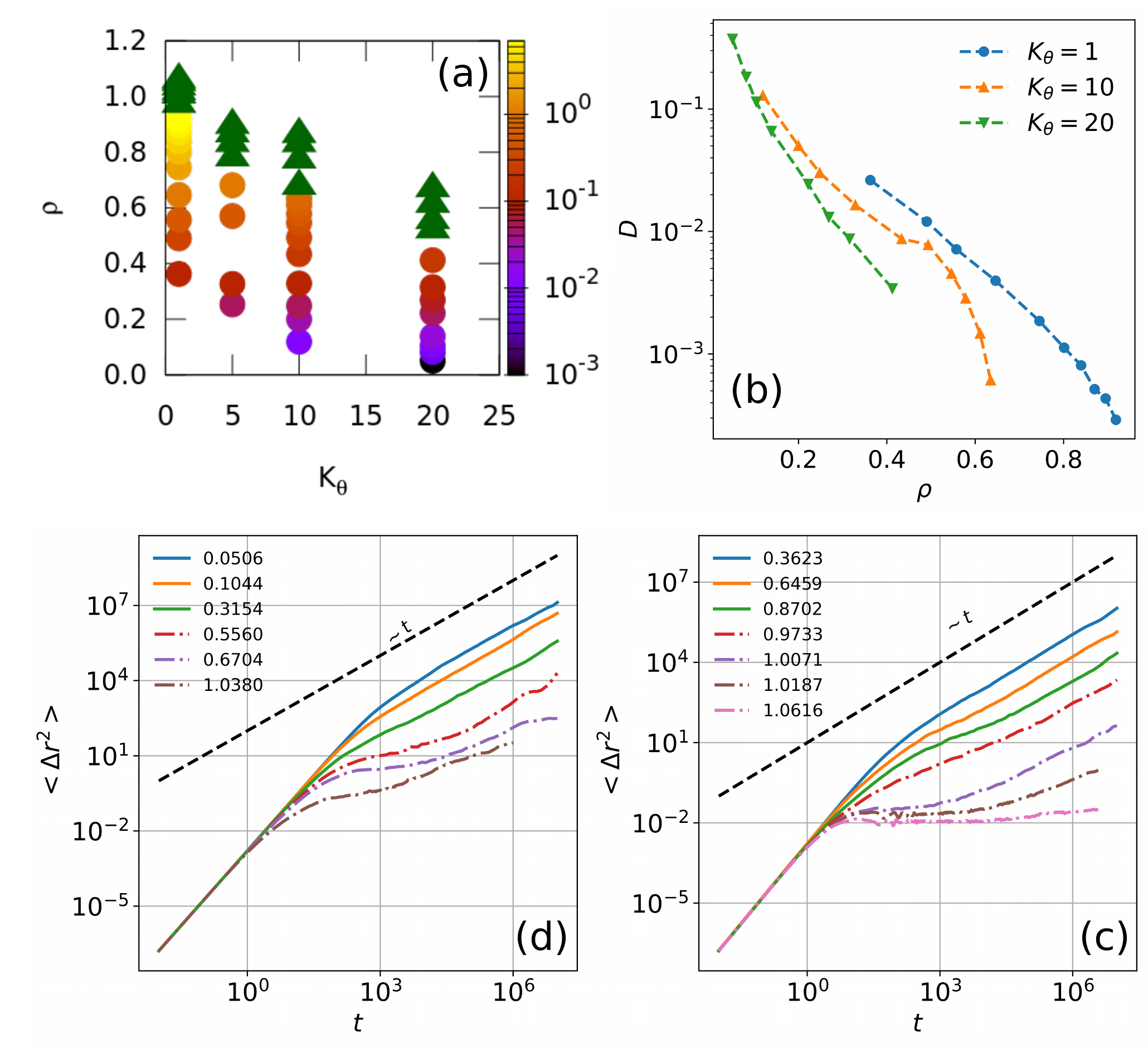}}
 \caption{(a) Average number density of monomers ($\rho$) vs angular stiffness of the rings ($K_{\theta}$), obtained for different values of pressure $P$ at $T=1$; the colorbar provides the scale of pressures for which equilibrium dynamics was observed. Data in green triangles correspond to regime where the dynamics is out-of-equilibrium within observation time-window.  Mean squared displacements of the centers of mass of the rings, $\langle\Delta{r^2}\rangle$, at different values of $\rho$ (listed in the keys) for $K_{\theta}=1$ (c) and $20$ (d); data in filled lines correspond to equilibrium conditions and data in dashed lines correspond to out-of-equilibrium conditions. (b) Variation of diffusion coefficient, $D$, with density $\rho$ for different $K_{\theta}$ values.}
\label{fig:1}
\end{figure}

\begin{figure}[t]
\centerline{\includegraphics[width=1.0\columnwidth]{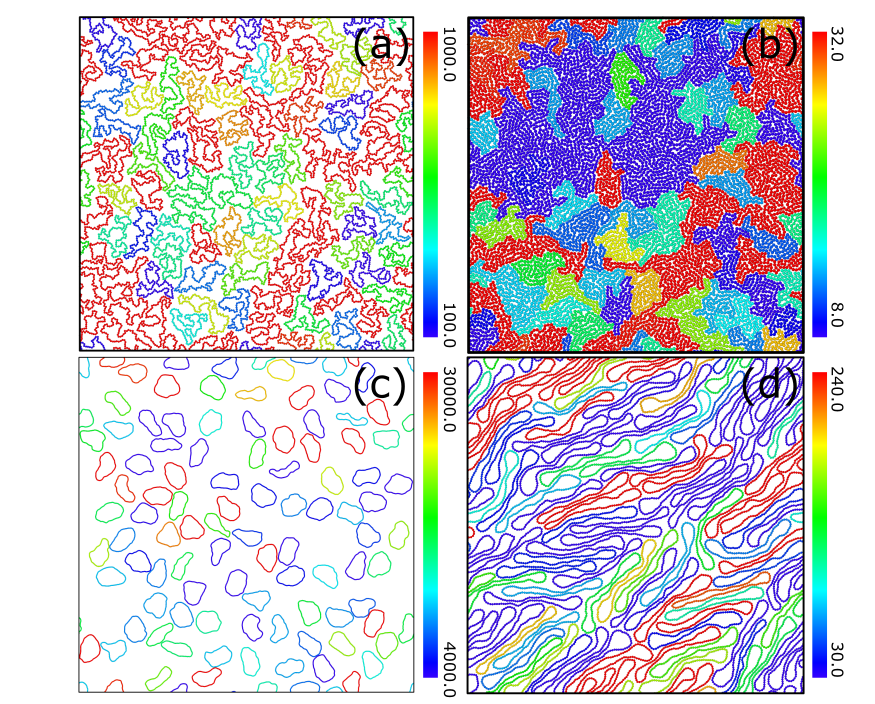}}
\caption{Snapshots of the ring polymer assembly at different equilibrium state points: 
[Top] $K_{\theta}=1$. (a) $\rho=0.3623$; (b) $\rho=0.9182$. [Bottom] $K_{\theta}=20$.  (c) $\rho=0.0506$; (d) $\rho=0.4122$. The rings are coloured as per respective squared displacement measured over time ${\Delta}t=10^4$, the magnitude of which is represented by the adjoining colorbars.}
\label{fig:2}
\end{figure}


In Figure \ref{fig:1}(a), we present a phase diagram that illustrates the range of equilibrated systems achievable within our simulation time window, based on the values of ring stiffness $K_\theta$ and density ($\rho$). The corresponding pressures are indicated by the colorbar. Additionally, we mark the region (indicated by green triangles) in the $K_\theta$-$\rho$ domain where the system remains out of equilibrium within our observation window. This region tentatively represents the glassy regime for these systems in the phase space. The phase diagram reveals that highly flexible rings allow for higher accessible densities within the equilibrium regime. Consequently, the resulting systems exhibit packings of crumpled rings, as shown in Figure \ref{fig:2}(b). As the chain stiffness increases, the achievable densities decrease, and the polymers exhibit less crumpled structures but rather deformed ring configurations. For the highest values of ring stiffness studied ($K_\theta=20$), the densely packed structures exhibit prolate ring structures, as depicted in Figure \ref{fig:2}(d). It is important to note that in previous simulations by Gnan et al. \cite{gnan2021dynamical}, much smaller rings were investigated. Thus, some of the deformed conformations observed in our study are solely due to the ring size, highlighting the influence of ring size on emergent properties in highly dense 2D systems.

In Figure \ref{fig:1}(b), we present the diffusion constants obtained from the long-time behavior of the center-of-mass mean square displacements (MSDs) for all the systems indicated in the equilibrium phase diagram shown in Figure \ref{fig:1}(a). The center-of-mass MSD data for flexible ($K_\theta=1$) and stiffer ($K_\theta=20$) polymer rings are depicted in Figures \ref{fig:1}(c) and \ref{fig:1}(d), respectively. These MSDs are computed using the NVT ensemble after equilibrating the system at the average density corresponding to the applied pressure, which is achieved using the NPT ensemble. The MSD data exhibits an initial ballistic regime at early times, followed by a diffusive regime at later times when the system reaches equilibrium dynamics. The diffusion coefficient ($D$) data demonstrates that all systems experience a slowing down with increasing density, characterized by decreasing diffusion constants, consistent with previous studies \cite{gnan2019microscopic, gnan2021dynamical}. Importantly, we observe the emergence of dynamically arrested states at lower densities as the ring stiffness increases, which aligns with findings in three-dimensional ring polymer assemblies \cite{roy2022effect}. It is crucial to note that the underlying mechanisms in two-dimensional systems are distinct from those in three-dimensional systems, as entanglements through threadings are not possible in two dimensions. Instead, the effect of crowding due to increasing density, combined with the role of ring shapes, drives densification and subsequent slowing down in our study. The data clearly demonstrates that, at a specific density $\rho$, stiffer ring polymers exhibit slower dynamics compared to flexible rings. Furthermore, the stiffer ring assemblies can even become dynamically arrested within a density range, whereas the assemblies of flexible rings remain mobile. Note that for $K=10$, we observe slight non-monotonicity in the density dependence of the diffusion coefficient, which could be a signature of the re-entrant behaviour reported earlier \cite{gnan2019microscopic, gnan2021dynamical}.

In Figures \ref{fig:1}(c) and \ref{fig:1}(d), we also include the MSD data (indicated by dashed lines) for higher pressures where equilibrium dynamics is not achieved within our observation time window. These data points are also marked in the phase diagram shown in Figure \ref{fig:1}(a), alongside similar data points for other values of $K_\theta$ that we have investigated. The MSD curves exhibit prolonged plateaus and late-time sub-diffusion, indicating the presence of caging effects imposed by neighboring rings and subsequent breaking of these cages for a subset of rings. These features are characteristic of aging glassy systems. In the next section, we delve into the role of density in relation to the deformation of ring polymers. We investigate how changes in the shapes of ring polymers are associated with local structural ordering and how this relates to the emergence of slow dynamics.

\subsection{Effect of stiffness on structural properties}

\subsubsection{Shape changes}

\begin{figure}[]
\centering
\includegraphics[width=1.0\columnwidth]{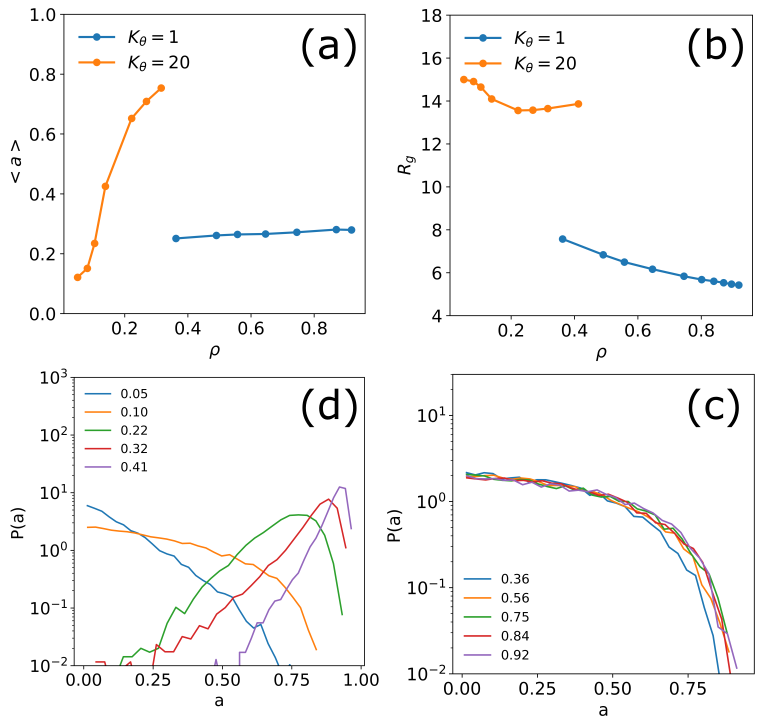}\\
\caption{(Top) For $K_{\theta}=1, 20$, variation with density  of (a) average asphericity $\langle{a}\rangle$ and (b) average radius of gyration of the rings $R_g$. (Bottom) Evolution of distribution of asphericity, $P(a)$, for $K_{\theta}=1$ (c), $20$ (d), with density $\rho$ (listed via keys in each case).}
\label{fig:3}
\end{figure}

The snapshots presented in Figure~\ref{fig:2} clearly demonstrate that the shapes of the polymers undergo significant changes with variations in both density and ring stiffness. These changes have a profound impact on the thermal packing of ring polymers in two dimensions, which differs markedly from their behavior in three-dimensional systems where rings can cross each other or undergo threading. To emphasize the contrasting behavior observed at high and low flexibility, we will focus our discussion on the two extreme cases we have studied: $K_{\theta}=1$ and $K_{\theta}=20$. These cases represent the limits of flexibility and stiffness, respectively, and allow us to elucidate the distinct characteristics exhibited by the ring polymers in our study. To quantify these changes, we compute the asphericity of each ring and then average over all rings across all configurations sampled in the equilibrium regime,  for $K_{\theta}=1, 20$. The variation of ensemble averaged asphericity with density is shown in the Fig.\ref{fig:3}(a) -- it increases with density substantially for stiffer rings and remains largely unaffected by increase in density for flexible rings \footnote{We note that while average asphericity increases with the ring stiffness, there is a reversal at higher values of $K_{\theta}$ (data not shown) when the rings become circular in shape and the asphericity then nearly vanishes.}. 

In the case of highly flexible rings ($K_\theta=1$), the increased packing density and thermal fluctuations result in crumpled ring structures with low asphericity, indicating a nearly spherical shape. This behavior remains unchanged even at high densities, as the rings become more crumpled but still maintain their overall spherical shape. Consequently, the asphericity shows little sensitivity to density variations. However, the average radius of gyration ($R_g$) of the rings decreases with increasing density, indicating a higher degree of compaction as the crowding increases (Figure \ref{fig:3}(b)). For the case of $K_\theta=20$, the rings are initially more elongated due to the stiffness preventing crumpling at low density. As the density increases, the rings undergo compaction and transition from circular to elongated shapes (as shown in Figure \ref{fig:2}(c)-(d)). This is reflected in the decrease in the average radius of gyration ($R_g$) with increasing density. However, the more significant feature is the sharp variation of asphericity with density, indicating a drastic change in the shapes of the rings. At the highest density accessible in equilibrium in our simulations, the asphericity approaches a value close to 1, indicating highly elongated structures, which is visually confirmed by the snapshot in Figure \ref{fig:2}(d). 

A more detailed understanding of how asphericity changes with density and stiffness can be obtained by examining the distributions of single ring asphericity, denoted as $P(a)$. In Figure \ref{fig:3}(c)-(d), we show the distributions for $K_\theta=1$ and $K_\theta=20$ respectively. For the case of flexible rings (low stiffness), where the rings are crumpled, the shape of the distributions does not appear to change significantly with density. This is also evident from the time series of asphericity for a tagged ring (see Figure S.2 in the Supplementary Information), where random fluctuations around similar mean values are observed for both low and high densities. In contrast, for the stiffer rings, there is a distinct behavior. At low density, there is a higher likelihood of observing rings with smaller asphericity, although shapes with larger asphericity are also possible due to thermal fluctuations and interactions with neighboring rings. However, a drastic change occurs with increasing density ($\rho > 0.1$). The peak of the distribution shifts to higher asphericity values, consistent with the elongated shapes observed in Figure \ref{fig:2}(d), and the skewness shifts towards the left, indicating that it becomes less common to sample rings with low asphericity. This change in distribution is also evident from the time series of a tagged ring (see Figure S.4), where intermittent large spikes in asphericity are observed. Overall, these observations strongly suggest that there may be a density threshold beyond which compaction leads to a significant change in shape, resulting in a shift of $P(a)$ towards higher asphericities. The observed distributions are consistent with the findings of Gnan et al. \cite{gnan2021dynamical}. However, it should be noted that in our case, we observe larger asphericities due to the use of stiffer and larger rings in our simulations.
\begin{figure*}[]
  \centering
  \includegraphics[width=1.8\columnwidth]{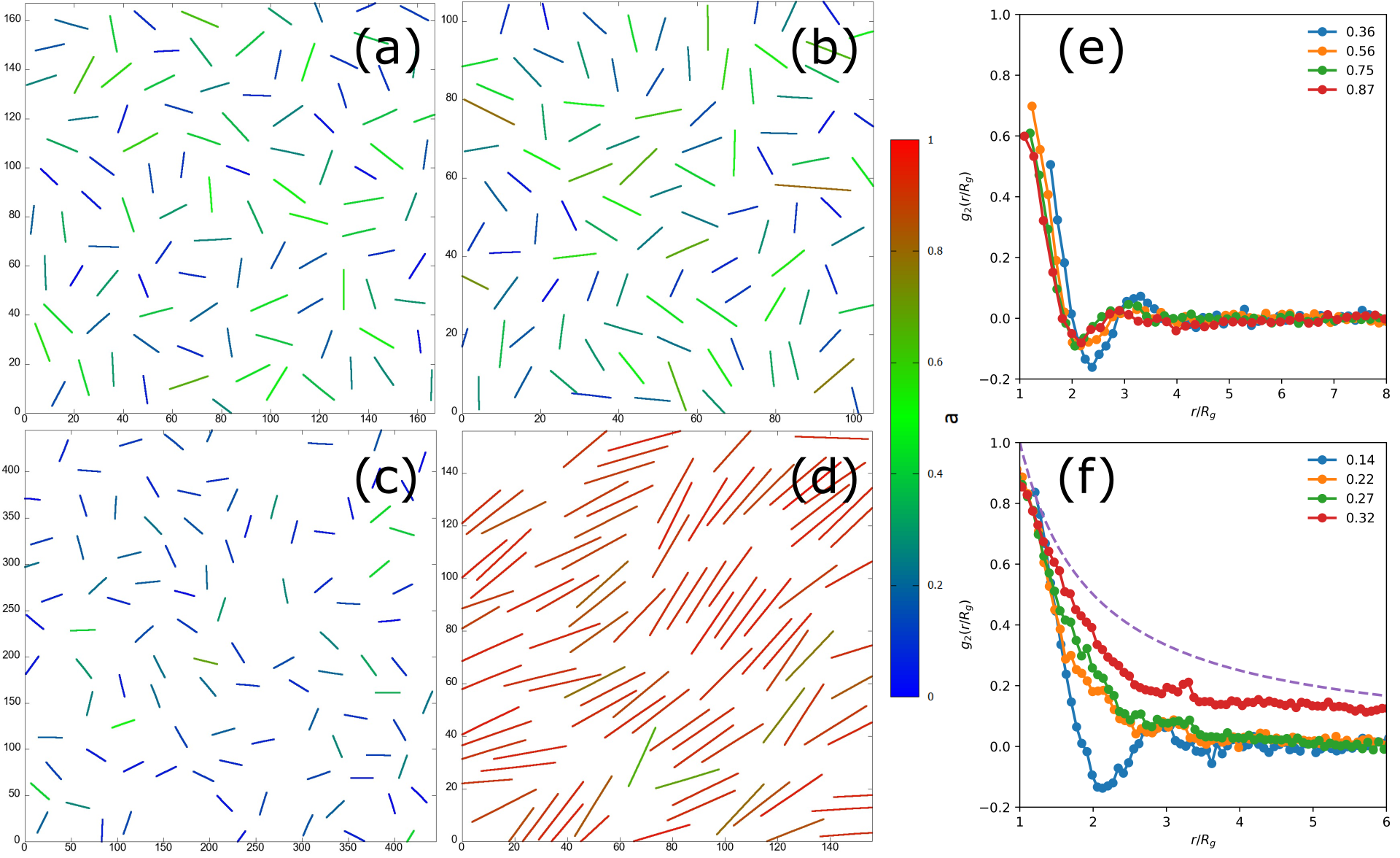}
 \caption{Probing  spatial ordering of the rings upon increasing $\rho$ and $K_{\theta}$.  (a)-(d). Rings represented via rods; centre of the rod located at the COM of the ring. The inclination of each rod represents the orientation of the large axis of  the ring, and the length of the rod represents the radius of gyration of the ring. The color of each rod represents the asphericity of the ring, $a$, as indicated by the colorbar. (a) $K_{\theta}=1$, $\rho=0.3623$; (b) $K_{\theta}=1$, $\rho=0.8702$; (c) $K_{\theta}=20$, $\rho=0.0814$; (d) $K_{\theta}=20$, $\rho=0.4122$. (e)-(f) Change in  orientational order, quantified by the orientational correlation function $g_{2}(r)$, with changing density (as marked) for $K_{\theta}=1$ (e) and $20$ (f). {The dashed line in (f) indicates a power-law with exponent $-1$.} 
 }
\label{fig:4}
\end{figure*}

\subsubsection{Spatial correlations}

In order to analyze and quantify the possible structural correlations in the system, we employ a representation where each ring is represented by a rod. The center of the rod is located at the center of mass of the ring, and its orientation is determined by the largest eigenvector of the gyration tensor of the ring. The length of the rod is chosen to be proportional to the radius of gyration, and the color of the rods is based on the measured asphericity of the corresponding ring. Figure \ref{fig:4}(a)-(d) shows illustrative snapshots using this representation for $K_\theta=1$ and $K_\theta=20$ at both low and high densities. For the case of $K_\theta=1$, the packing of rods and their orientations appear random at both low and high densities. However, for $K_\theta=20$, while the low density structure remains random, the high density structure exhibits spatial patches of aligned rods, indicating a degree of order in the packing of stiff rings in the most dense equilibrium structures achieved. Figure \ref{fig:4}(d) provides a visual representation of this spatial organization. To quantitatively analyze the extent of orientational order, we employ the orientational correlation function $g_2(r)$, which is defined in Methods section. This correlation function allows us to measure the correlations in ring orientations as a function of distance $r$.

In Fig.\ref{fig:4}(e)-(f), we present the variation of the orientational correlation function $g_2(r)$ with density for $K_\theta=1$ and $K_\theta=20$. In a nematic phase, $g_2(r)$ is expected to exhibit a power-law decay of the form $g_2(r) \propto r^{-\eta_2}$. For the case of $K_\theta=1$, $g_2(r)$ decays rapidly and shows anti-correlations over short distances for all explored densities. This indicates the absence of any significant orientational ordering in the system, regardless of density. For the case of $K_\theta=20$ at low density, $g_2(r)$ exhibits a similar behavior to that of $K_\theta=1$, with a quick decay and anti-correlations over short distances. However, at higher densities, we do not observe significant anti-correlations. This regime corresponds to highly aspherical ring packings, indicating a distinct behavior compared to both low density and flexible ring systems. For the highest density where equilibrated states can be sampled, we observe signatures of power-law correlations in $g_2(r)$. Furthermore, we attempt to extract a correlation length ($\xi_{\rm sr}$) from the short-range decay of $g_2(r)$, which appears to be exponential in nature. The calculated correlation lengths are $\xi_{\rm sr}=0.65687$, $0.796795$, and $1.62262$ for density values $\rho=0.22$, $0.27$, and $0.32$, respectively. These results indicate an increasing spatial correlation with increasing density, suggesting the emergence of orientational order in the system.

Finally, we try to relate the observed structural properties with the measured dynamics. In Fig.\ref{fig:2}, we have color-coded the rings based on their squared displacement within a fixed time period ($\Delta t=10^4$), focusing on the behavior at high densities. For highly flexible polymers (low $K_\theta$), which form crumpled globular structures without spatial ordering upon compaction, we observe a random spatial distribution of mobilities. This indicates the presence of dynamical heterogeneity, a characteristic feature of glass-forming systems \cite{berthier2011dynamic}. In our case, this heterogeneity arises from the motion of extended deformable objects. For the stiffer polymers (high $K_\theta$), where spatial correlations in ring orientations are observed, we can see that the regions having spatial order, also have reduced mobility. In other words, the growing spatial correlations resulting from specific shape changes upon densification are associated with the observed dynamical slowing down of the stiff rings. This onset of glassiness, characterized by medium-range order and the concomitant slowing down, is reminiscent of other glass-forming systems that display similar features \cite{kawasaki2007correlation}. Overall, the relationship between structural properties, such as spatial correlations and ring shapes, and the dynamics of the system suggests that the observed glassy behavior is an interplay of local structural arrangements and global ring deformations.

\section{Discussion and Conclusion}\label{SEC-IV}
In summary, motivated by recent explorations of glassy dynamics or jamming in two dimensional systems consisting of deformable soft objects, we have studied how dynamical behaviour of a 2D thermal assembly of ring polymers is affected via the tuning of the ring stiffness. 

Using extensive molecular dynamics simulations, our investigation revealed several key findings:
(1) dynamical slow-down: increasing pressure leads to dynamical slowing down, for all magnitudes of ring stiffness.
(2) phase diagram:  we mapped out the observed region of equilibrium behavior in the phase space of ring stiffness and average monomer density obtained from our NPT simulations; we observe onset of glassy behaviour  at smaller densities for stiffer polymers. 
(3) local structural ordering: at high pressures and ring stiffness, we demonstrate the emergence of spatial domains of orientational ordering.
(4) deformability and relation to slower dynamics: for flexible rings, the crowding drives the dynamical slowing down where as for stiffer rings, emergence of local ordered domains to a different length scale, coupled with crowding, underpins the slower dynamics. 
(5) 2D vs 3D: unlike three-dimensional systems where threading of inter-rings is a leading cause of glassy dynamics, in two-dimensional systems, the deformation of flexible rings was proposed as a viable mechanism for achieving slower dynamics and kinetically arrested states.
These findings highlight the role of the interplay of ring flexibility and crowding in determining the glassy dynamics of two-dimensional packings of ring polymers at finite temperature. 

Finally, we note that our work has focused on a monodisperse collection of ring polymers, in terms of size as well as stiffness. It would be interesting to investigate how increasing dispersity in size or stiffness affects our observations, especially in the context of the spatial ordering and consequent dynamics that we are observing. 

\section*{Acknowledgment}
 We thank the HPC facility at the Institute of Mathematical Sciences for computing time. PC and SV also acknowledge support via the sub-project on the Modeling of Soft Materials within the IMSc Apex project, funded by Dept. of Atomic Energy, Government of India.

\section*{Data Availability}
The data that support the findings of this study are available from the corresponding author upon reasonable request. 

\bibliography{2D}

\begin{thebibliography}{30}%
\makeatletter
\providecommand \@ifxundefined [1]{%
 \@ifx{#1\undefined}
}%
\providecommand \@ifnum [1]{%
 \ifnum #1\expandafter \@firstoftwo
 \else \expandafter \@secondoftwo
 \fi
}%
\providecommand \@ifx [1]{%
 \ifx #1\expandafter \@firstoftwo
 \else \expandafter \@secondoftwo
 \fi
}%
\providecommand \natexlab [1]{#1}%
\providecommand \enquote  [1]{``#1''}%
\providecommand \bibnamefont  [1]{#1}%
\providecommand \bibfnamefont [1]{#1}%
\providecommand \citenamefont [1]{#1}%
\providecommand \href@noop [0]{\@secondoftwo}%
\providecommand \href [0]{\begingroup \@sanitize@url \@href}%
\providecommand \@href[1]{\@@startlink{#1}\@@href}%
\providecommand \@@href[1]{\endgroup#1\@@endlink}%
\providecommand \@sanitize@url [0]{\catcode `\\12\catcode `\$12\catcode
  `\&12\catcode `\#12\catcode `\^12\catcode `\_12\catcode `\%12\relax}%
\providecommand \@@startlink[1]{}%
\providecommand \@@endlink[0]{}%
\providecommand \url  [0]{\begingroup\@sanitize@url \@url }%
\providecommand \@url [1]{\endgroup\@href {#1}{\urlprefix }}%
\providecommand \urlprefix  [0]{URL }%
\providecommand \Eprint [0]{\href }%
\providecommand \doibase [0]{http://dx.doi.org/}%
\providecommand \selectlanguage [0]{\@gobble}%
\providecommand \bibinfo  [0]{\@secondoftwo}%
\providecommand \bibfield  [0]{\@secondoftwo}%
\providecommand \translation [1]{[#1]}%
\providecommand \BibitemOpen [0]{}%
\providecommand \bibitemStop [0]{}%
\providecommand \bibitemNoStop [0]{.\EOS\space}%
\providecommand \EOS [0]{\spacefactor3000\relax}%
\providecommand \BibitemShut  [1]{\csname bibitem#1\endcsname}%
\let\auto@bib@innerbib\@empty
\bibitem [{\citenamefont {Binder}\ and\ \citenamefont
  {Kob}(2011)}]{binder2011glassy}%
  \BibitemOpen
  \bibfield  {author} {\bibinfo {author} {\bibfnamefont {K.}~\bibnamefont
  {Binder}}\ and\ \bibinfo {author} {\bibfnamefont {W.}~\bibnamefont {Kob}},\
  }\href@noop {} {\emph {\bibinfo {title} {Glassy materials and disordered
  solids: An introduction to their statistical mechanics}}}\ (\bibinfo
  {publisher} {World scientific},\ \bibinfo {year} {2011})\BibitemShut
  {NoStop}%
\bibitem [{\citenamefont {Parisi}\ and\ \citenamefont
  {Zamponi}(2010)}]{parisi2010mean}%
  \BibitemOpen
  \bibfield  {author} {\bibinfo {author} {\bibfnamefont {G.}~\bibnamefont
  {Parisi}}\ and\ \bibinfo {author} {\bibfnamefont {F.}~\bibnamefont
  {Zamponi}},\ }\bibfield  {title} {\enquote {\bibinfo {title} {Mean-field
  theory of hard sphere glasses and jamming},}\ }\href@noop {} {\bibfield
  {journal} {\bibinfo  {journal} {Reviews of Modern Physics}\ }\textbf
  {\bibinfo {volume} {82}},\ \bibinfo {pages} {789} (\bibinfo {year}
  {2010})}\BibitemShut {NoStop}%
\bibitem [{\citenamefont {Mukhopadhyay}\ and\ \citenamefont
  {Peixinho}(2011)}]{mukhopadhyay2011packings}%
  \BibitemOpen
  \bibfield  {author} {\bibinfo {author} {\bibfnamefont {S.}~\bibnamefont
  {Mukhopadhyay}}\ and\ \bibinfo {author} {\bibfnamefont {J.}~\bibnamefont
  {Peixinho}},\ }\bibfield  {title} {\enquote {\bibinfo {title} {Packings of
  deformable spheres},}\ }\href@noop {} {\bibfield  {journal} {\bibinfo
  {journal} {Physical Review E}\ }\textbf {\bibinfo {volume} {84}},\ \bibinfo
  {pages} {011302} (\bibinfo {year} {2011})}\BibitemShut {NoStop}%
\bibitem [{\citenamefont {Makse}, \citenamefont {Johnson},\ and\ \citenamefont
  {Schwartz}(2000)}]{makse2000packing}%
  \BibitemOpen
  \bibfield  {author} {\bibinfo {author} {\bibfnamefont {H.~A.}\ \bibnamefont
  {Makse}}, \bibinfo {author} {\bibfnamefont {D.~L.}\ \bibnamefont {Johnson}},
  \ and\ \bibinfo {author} {\bibfnamefont {L.~M.}\ \bibnamefont {Schwartz}},\
  }\bibfield  {title} {\enquote {\bibinfo {title} {Packing of compressible
  granular materials},}\ }\href@noop {} {\bibfield  {journal} {\bibinfo
  {journal} {Physical review letters}\ }\textbf {\bibinfo {volume} {84}},\
  \bibinfo {pages} {4160} (\bibinfo {year} {2000})}\BibitemShut {NoStop}%
\bibitem [{\citenamefont {Batista}\ and\ \citenamefont
  {Miller}(2010)}]{batista2010crystallization}%
  \BibitemOpen
  \bibfield  {author} {\bibinfo {author} {\bibfnamefont {V.~M.}\ \bibnamefont
  {Batista}}\ and\ \bibinfo {author} {\bibfnamefont {M.~A.}\ \bibnamefont
  {Miller}},\ }\bibfield  {title} {\enquote {\bibinfo {title} {Crystallization
  of deformable spherical colloids},}\ }\href@noop {} {\bibfield  {journal}
  {\bibinfo  {journal} {Physical review letters}\ }\textbf {\bibinfo {volume}
  {105}},\ \bibinfo {pages} {088305} (\bibinfo {year} {2010})}\BibitemShut
  {NoStop}%
\bibitem [{\citenamefont {Boromand}\ \emph {et~al.}(2018)\citenamefont
  {Boromand}, \citenamefont {Signoriello}, \citenamefont {Ye}, \citenamefont
  {O’Hern},\ and\ \citenamefont {Shattuck}}]{boromand2018jamming}%
  \BibitemOpen
  \bibfield  {author} {\bibinfo {author} {\bibfnamefont {A.}~\bibnamefont
  {Boromand}}, \bibinfo {author} {\bibfnamefont {A.}~\bibnamefont
  {Signoriello}}, \bibinfo {author} {\bibfnamefont {F.}~\bibnamefont {Ye}},
  \bibinfo {author} {\bibfnamefont {C.~S.}\ \bibnamefont {O’Hern}}, \ and\
  \bibinfo {author} {\bibfnamefont {M.~D.}\ \bibnamefont {Shattuck}},\
  }\bibfield  {title} {\enquote {\bibinfo {title} {Jamming of deformable
  polygons},}\ }\href@noop {} {\bibfield  {journal} {\bibinfo  {journal}
  {Physical review letters}\ }\textbf {\bibinfo {volume} {121}},\ \bibinfo
  {pages} {248003} (\bibinfo {year} {2018})}\BibitemShut {NoStop}%
\bibitem [{\citenamefont {Gnan}\ and\ \citenamefont
  {Zaccarelli}(2019)}]{gnan2019microscopic}%
  \BibitemOpen
  \bibfield  {author} {\bibinfo {author} {\bibfnamefont {N.}~\bibnamefont
  {Gnan}}\ and\ \bibinfo {author} {\bibfnamefont {E.}~\bibnamefont
  {Zaccarelli}},\ }\bibfield  {title} {\enquote {\bibinfo {title} {The
  microscopic role of deformation in the dynamics of soft colloids},}\
  }\href@noop {} {\bibfield  {journal} {\bibinfo  {journal} {Nature Physics}\
  }\textbf {\bibinfo {volume} {15}},\ \bibinfo {pages} {683--688} (\bibinfo
  {year} {2019})}\BibitemShut {NoStop}%
\bibitem [{\citenamefont {Ge}, \citenamefont {Panyukov},\ and\ \citenamefont
  {Rubinstein}(2016)}]{ge2016self}%
  \BibitemOpen
  \bibfield  {author} {\bibinfo {author} {\bibfnamefont {T.}~\bibnamefont
  {Ge}}, \bibinfo {author} {\bibfnamefont {S.}~\bibnamefont {Panyukov}}, \ and\
  \bibinfo {author} {\bibfnamefont {M.}~\bibnamefont {Rubinstein}},\ }\bibfield
   {title} {\enquote {\bibinfo {title} {Self-similar conformations and dynamics
  in entangled melts and solutions of nonconcatenated ring polymers},}\
  }\href@noop {} {\bibfield  {journal} {\bibinfo  {journal} {Macromolecules}\
  }\textbf {\bibinfo {volume} {49}},\ \bibinfo {pages} {708--722} (\bibinfo
  {year} {2016})}\BibitemShut {NoStop}%
\bibitem [{\citenamefont {Halverson}\ \emph {et~al.}(2014)\citenamefont
  {Halverson}, \citenamefont {Smrek}, \citenamefont {Kremer},\ and\
  \citenamefont {Grosberg}}]{halverson2014melt}%
  \BibitemOpen
  \bibfield  {author} {\bibinfo {author} {\bibfnamefont {J.~D.}\ \bibnamefont
  {Halverson}}, \bibinfo {author} {\bibfnamefont {J.}~\bibnamefont {Smrek}},
  \bibinfo {author} {\bibfnamefont {K.}~\bibnamefont {Kremer}}, \ and\ \bibinfo
  {author} {\bibfnamefont {A.~Y.}\ \bibnamefont {Grosberg}},\ }\bibfield
  {title} {\enquote {\bibinfo {title} {From a melt of rings to chromosome
  territories: the role of topological constraints in genome folding},}\
  }\href@noop {} {\bibfield  {journal} {\bibinfo  {journal} {Reports on
  Progress in Physics}\ }\textbf {\bibinfo {volume} {77}},\ \bibinfo {pages}
  {022601} (\bibinfo {year} {2014})}\BibitemShut {NoStop}%
\bibitem [{\citenamefont {Michieletto}\ and\ \citenamefont
  {Turner}(2016)}]{michieletto2016topologically}%
  \BibitemOpen
  \bibfield  {author} {\bibinfo {author} {\bibfnamefont {D.}~\bibnamefont
  {Michieletto}}\ and\ \bibinfo {author} {\bibfnamefont {M.~S.}\ \bibnamefont
  {Turner}},\ }\bibfield  {title} {\enquote {\bibinfo {title} {A topologically
  driven glass in ring polymers},}\ }\href@noop {} {\bibfield  {journal}
  {\bibinfo  {journal} {Proceedings of the National Academy of Sciences}\
  }\textbf {\bibinfo {volume} {113}},\ \bibinfo {pages} {5195--5200} (\bibinfo
  {year} {2016})}\BibitemShut {NoStop}%
\bibitem [{\citenamefont {Michieletto}, \citenamefont {Nahali},\ and\
  \citenamefont {Rosa}(2017)}]{michieletto2017glassiness}%
  \BibitemOpen
  \bibfield  {author} {\bibinfo {author} {\bibfnamefont {D.}~\bibnamefont
  {Michieletto}}, \bibinfo {author} {\bibfnamefont {N.}~\bibnamefont {Nahali}},
  \ and\ \bibinfo {author} {\bibfnamefont {A.}~\bibnamefont {Rosa}},\
  }\bibfield  {title} {\enquote {\bibinfo {title} {Glassiness and heterogeneous
  dynamics in dense solutions of ring polymers},}\ }\href@noop {} {\bibfield
  {journal} {\bibinfo  {journal} {Physical Review Letters}\ }\textbf {\bibinfo
  {volume} {119}},\ \bibinfo {pages} {197801} (\bibinfo {year}
  {2017})}\BibitemShut {NoStop}%
\bibitem [{\citenamefont {Louis}\ \emph {et~al.}(2000)\citenamefont {Louis},
  \citenamefont {Bolhuis}, \citenamefont {Hansen},\ and\ \citenamefont
  {Meijer}}]{louis2000can}%
  \BibitemOpen
  \bibfield  {author} {\bibinfo {author} {\bibfnamefont {A.}~\bibnamefont
  {Louis}}, \bibinfo {author} {\bibfnamefont {P.}~\bibnamefont {Bolhuis}},
  \bibinfo {author} {\bibfnamefont {J.}~\bibnamefont {Hansen}}, \ and\ \bibinfo
  {author} {\bibfnamefont {E.}~\bibnamefont {Meijer}},\ }\bibfield  {title}
  {\enquote {\bibinfo {title} {Can polymer coils be modeled as “soft
  colloids”?}}\ }\href@noop {} {\bibfield  {journal} {\bibinfo  {journal}
  {Physical review letters}\ }\textbf {\bibinfo {volume} {85}},\ \bibinfo
  {pages} {2522} (\bibinfo {year} {2000})}\BibitemShut {NoStop}%
\bibitem [{\citenamefont {G{\"o}tze}, \citenamefont {Harreis},\ and\
  \citenamefont {Likos}(2004)}]{gotze2004tunable}%
  \BibitemOpen
  \bibfield  {author} {\bibinfo {author} {\bibfnamefont {I.}~\bibnamefont
  {G{\"o}tze}}, \bibinfo {author} {\bibfnamefont {H.}~\bibnamefont {Harreis}},
  \ and\ \bibinfo {author} {\bibfnamefont {C.}~\bibnamefont {Likos}},\
  }\bibfield  {title} {\enquote {\bibinfo {title} {Tunable effective
  interactions between dendritic macromolecules},}\ }\href@noop {} {\bibfield
  {journal} {\bibinfo  {journal} {The Journal of chemical physics}\ }\textbf
  {\bibinfo {volume} {120}},\ \bibinfo {pages} {7761--7771} (\bibinfo {year}
  {2004})}\BibitemShut {NoStop}%
\bibitem [{\citenamefont {O?hern}\ \emph {et~al.}(2003)\citenamefont {O?hern},
  \citenamefont {Silbert}, \citenamefont {Liu},\ and\ \citenamefont
  {Nagel}}]{o2003jamming}%
  \BibitemOpen
  \bibfield  {author} {\bibinfo {author} {\bibfnamefont {C.~S.}\ \bibnamefont
  {O?hern}}, \bibinfo {author} {\bibfnamefont {L.~E.}\ \bibnamefont {Silbert}},
  \bibinfo {author} {\bibfnamefont {A.~J.}\ \bibnamefont {Liu}}, \ and\
  \bibinfo {author} {\bibfnamefont {S.~R.}\ \bibnamefont {Nagel}},\ }\bibfield
  {title} {\enquote {\bibinfo {title} {Jamming at zero temperature and zero
  applied stress: The epitome of disorder},}\ }\href@noop {} {\bibfield
  {journal} {\bibinfo  {journal} {Physical Review E}\ }\textbf {\bibinfo
  {volume} {68}},\ \bibinfo {pages} {011306} (\bibinfo {year}
  {2003})}\BibitemShut {NoStop}%
\bibitem [{\citenamefont {van Hecke}(2009)}]{van2009jamming}%
  \BibitemOpen
  \bibfield  {author} {\bibinfo {author} {\bibfnamefont {M.}~\bibnamefont {van
  Hecke}},\ }\bibfield  {title} {\enquote {\bibinfo {title} {Jamming of soft
  particles: geometry, mechanics, scaling and isostaticity},}\ }\href@noop {}
  {\bibfield  {journal} {\bibinfo  {journal} {Journal of Physics: Condensed
  Matter}\ }\textbf {\bibinfo {volume} {22}},\ \bibinfo {pages} {033101}
  (\bibinfo {year} {2009})}\BibitemShut {NoStop}%
\bibitem [{\citenamefont {Chaudhuri}, \citenamefont {Berthier},\ and\
  \citenamefont {Sastry}(2010)}]{chaudhuri2010jamming}%
  \BibitemOpen
  \bibfield  {author} {\bibinfo {author} {\bibfnamefont {P.}~\bibnamefont
  {Chaudhuri}}, \bibinfo {author} {\bibfnamefont {L.}~\bibnamefont {Berthier}},
  \ and\ \bibinfo {author} {\bibfnamefont {S.}~\bibnamefont {Sastry}},\
  }\bibfield  {title} {\enquote {\bibinfo {title} {Jamming transitions in
  amorphous packings of frictionless spheres occur over a continuous range of
  volume fractions},}\ }\href@noop {} {\bibfield  {journal} {\bibinfo
  {journal} {Physical review letters}\ }\textbf {\bibinfo {volume} {104}},\
  \bibinfo {pages} {165701} (\bibinfo {year} {2010})}\BibitemShut {NoStop}%
\bibitem [{\citenamefont {Ikeda}, \citenamefont {Berthier},\ and\ \citenamefont
  {Sollich}(2012)}]{ikeda2012unified}%
  \BibitemOpen
  \bibfield  {author} {\bibinfo {author} {\bibfnamefont {A.}~\bibnamefont
  {Ikeda}}, \bibinfo {author} {\bibfnamefont {L.}~\bibnamefont {Berthier}}, \
  and\ \bibinfo {author} {\bibfnamefont {P.}~\bibnamefont {Sollich}},\
  }\bibfield  {title} {\enquote {\bibinfo {title} {Unified study of glass and
  jamming rheology in soft particle systems},}\ }\href@noop {} {\bibfield
  {journal} {\bibinfo  {journal} {Physical review letters}\ }\textbf {\bibinfo
  {volume} {109}},\ \bibinfo {pages} {018301} (\bibinfo {year}
  {2012})}\BibitemShut {NoStop}%
\bibitem [{\citenamefont {Manning}(2023)}]{manning2023essay}%
  \BibitemOpen
  \bibfield  {author} {\bibinfo {author} {\bibfnamefont {M.~L.}\ \bibnamefont
  {Manning}},\ }\bibfield  {title} {\enquote {\bibinfo {title} {Essay:
  Collections of deformable particles present exciting challenges for soft
  matter and biological physics},}\ }\href@noop {} {\bibfield  {journal}
  {\bibinfo  {journal} {Physical Review Letters}\ }\textbf {\bibinfo {volume}
  {130}},\ \bibinfo {pages} {130002} (\bibinfo {year} {2023})}\BibitemShut
  {NoStop}%
\bibitem [{\citenamefont {Loewe}\ \emph {et~al.}(2020)\citenamefont {Loewe},
  \citenamefont {Chiang}, \citenamefont {Marenduzzo},\ and\ \citenamefont
  {Marchetti}}]{Loewe2020}%
  \BibitemOpen
  \bibfield  {author} {\bibinfo {author} {\bibfnamefont {B.}~\bibnamefont
  {Loewe}}, \bibinfo {author} {\bibfnamefont {M.}~\bibnamefont {Chiang}},
  \bibinfo {author} {\bibfnamefont {D.}~\bibnamefont {Marenduzzo}}, \ and\
  \bibinfo {author} {\bibfnamefont {M.~C.}\ \bibnamefont {Marchetti}},\
  }\bibfield  {title} {\enquote {\bibinfo {title} {Solid-liquid transition of
  deformable and overlapping active particles},}\ }\href {\doibase
  10.1103/PhysRevLett.125.038003} {\bibfield  {journal} {\bibinfo  {journal}
  {Phys. Rev. Lett.}\ }\textbf {\bibinfo {volume} {125}},\ \bibinfo {pages}
  {038003} (\bibinfo {year} {2020})}\BibitemShut {NoStop}%
\bibitem [{\citenamefont {{\v{S}}iber}\ and\ \citenamefont
  {Ziherl}(2013)}]{vsiber2013many}%
  \BibitemOpen
  \bibfield  {author} {\bibinfo {author} {\bibfnamefont {A.}~\bibnamefont
  {{\v{S}}iber}}\ and\ \bibinfo {author} {\bibfnamefont {P.}~\bibnamefont
  {Ziherl}},\ }\bibfield  {title} {\enquote {\bibinfo {title} {Many-body
  contact repulsion of deformable disks},}\ }\href@noop {} {\bibfield
  {journal} {\bibinfo  {journal} {Physical review letters}\ }\textbf {\bibinfo
  {volume} {110}},\ \bibinfo {pages} {214301} (\bibinfo {year}
  {2013})}\BibitemShut {NoStop}%
\bibitem [{\citenamefont {Miller}\ and\ \citenamefont
  {Cacciuto}(2011)}]{miller2011two}%
  \BibitemOpen
  \bibfield  {author} {\bibinfo {author} {\bibfnamefont {W.~L.}\ \bibnamefont
  {Miller}}\ and\ \bibinfo {author} {\bibfnamefont {A.}~\bibnamefont
  {Cacciuto}},\ }\bibfield  {title} {\enquote {\bibinfo {title}
  {Two-dimensional packing of soft particles and the soft generalized thomson
  problem},}\ }\href@noop {} {\bibfield  {journal} {\bibinfo  {journal} {Soft
  Matter}\ }\textbf {\bibinfo {volume} {7}},\ \bibinfo {pages} {7552--7559}
  (\bibinfo {year} {2011})}\BibitemShut {NoStop}%
\bibitem [{\citenamefont {Treado}\ \emph {et~al.}(2021)\citenamefont {Treado},
  \citenamefont {Wang}, \citenamefont {Boromand}, \citenamefont {Murrell},
  \citenamefont {Shattuck},\ and\ \citenamefont {O'Hern}}]{treado2021bridging}%
  \BibitemOpen
  \bibfield  {author} {\bibinfo {author} {\bibfnamefont {J.~D.}\ \bibnamefont
  {Treado}}, \bibinfo {author} {\bibfnamefont {D.}~\bibnamefont {Wang}},
  \bibinfo {author} {\bibfnamefont {A.}~\bibnamefont {Boromand}}, \bibinfo
  {author} {\bibfnamefont {M.~P.}\ \bibnamefont {Murrell}}, \bibinfo {author}
  {\bibfnamefont {M.~D.}\ \bibnamefont {Shattuck}}, \ and\ \bibinfo {author}
  {\bibfnamefont {C.~S.}\ \bibnamefont {O'Hern}},\ }\bibfield  {title}
  {\enquote {\bibinfo {title} {Bridging particle deformability and collective
  response in soft solids},}\ }\href@noop {} {\bibfield  {journal} {\bibinfo
  {journal} {Physical Review Materials}\ }\textbf {\bibinfo {volume} {5}},\
  \bibinfo {pages} {055605} (\bibinfo {year} {2021})}\BibitemShut {NoStop}%
\bibitem [{\citenamefont {Smrek}\ \emph {et~al.}(2020)\citenamefont {Smrek},
  \citenamefont {Chubak}, \citenamefont {Likos},\ and\ \citenamefont
  {Kremer}}]{smrek2020active}%
  \BibitemOpen
  \bibfield  {author} {\bibinfo {author} {\bibfnamefont {J.}~\bibnamefont
  {Smrek}}, \bibinfo {author} {\bibfnamefont {I.}~\bibnamefont {Chubak}},
  \bibinfo {author} {\bibfnamefont {C.~N.}\ \bibnamefont {Likos}}, \ and\
  \bibinfo {author} {\bibfnamefont {K.}~\bibnamefont {Kremer}},\ }\bibfield
  {title} {\enquote {\bibinfo {title} {Active topological glass},}\ }\href@noop
  {} {\bibfield  {journal} {\bibinfo  {journal} {Nature Communications}\
  }\textbf {\bibinfo {volume} {11}} (\bibinfo {year} {2020})}\BibitemShut
  {NoStop}%
\bibitem [{\citenamefont {Plimpton}(1995)}]{plimpton1995fast}%
  \BibitemOpen
  \bibfield  {author} {\bibinfo {author} {\bibfnamefont {S.}~\bibnamefont
  {Plimpton}},\ }\bibfield  {title} {\enquote {\bibinfo {title} {Fast parallel
  algorithms for short-range molecular dynamics},}\ }\href@noop {} {\bibfield
  {journal} {\bibinfo  {journal} {Journal of computational physics}\ }\textbf
  {\bibinfo {volume} {117}},\ \bibinfo {pages} {1--19} (\bibinfo {year}
  {1995})}\BibitemShut {NoStop}%
\bibitem [{\citenamefont {Gnan}\ \emph {et~al.}(2021)\citenamefont {Gnan},
  \citenamefont {Camerin}, \citenamefont {Del~Monte}, \citenamefont
  {Ninarello},\ and\ \citenamefont {Zaccarelli}}]{gnan2021dynamical}%
  \BibitemOpen
  \bibfield  {author} {\bibinfo {author} {\bibfnamefont {N.}~\bibnamefont
  {Gnan}}, \bibinfo {author} {\bibfnamefont {F.}~\bibnamefont {Camerin}},
  \bibinfo {author} {\bibfnamefont {G.}~\bibnamefont {Del~Monte}}, \bibinfo
  {author} {\bibfnamefont {A.}~\bibnamefont {Ninarello}}, \ and\ \bibinfo
  {author} {\bibfnamefont {E.}~\bibnamefont {Zaccarelli}},\ }\bibfield  {title}
  {\enquote {\bibinfo {title} {Dynamical properties of different models of
  elastic polymer rings: Confirming the link between deformation and
  fragility},}\ }\href@noop {} {\bibfield  {journal} {\bibinfo  {journal} {The
  Journal of Chemical Physics}\ }\textbf {\bibinfo {volume} {154}},\ \bibinfo
  {pages} {154901} (\bibinfo {year} {2021})}\BibitemShut {NoStop}%
\bibitem [{\citenamefont {Bates}\ and\ \citenamefont
  {Frenkel}(2000)}]{Frenkel2000}%
  \BibitemOpen
  \bibfield  {author} {\bibinfo {author} {\bibfnamefont {M.~A.}\ \bibnamefont
  {Bates}}\ and\ \bibinfo {author} {\bibfnamefont {D.}~\bibnamefont
  {Frenkel}},\ }\bibfield  {title} {\enquote {\bibinfo {title} {Phase behavior
  of two-dimensional hard rod fluids},}\ }\href {\doibase 10.1063/1.481637}
  {\bibfield  {journal} {\bibinfo  {journal} {The Journal of Chemical Physics}\
  }\textbf {\bibinfo {volume} {112}},\ \bibinfo {pages} {10034--10041}
  (\bibinfo {year} {2000})},\ \Eprint
  {http://arxiv.org/abs/https://doi.org/10.1063/1.481637}
  {https://doi.org/10.1063/1.481637} \BibitemShut {NoStop}%
\bibitem [{\citenamefont {Roy}, \citenamefont {Chaudhuri},\ and\ \citenamefont
  {Vemparala}(2022)}]{roy2022effect}%
  \BibitemOpen
  \bibfield  {author} {\bibinfo {author} {\bibfnamefont {P.~K.}\ \bibnamefont
  {Roy}}, \bibinfo {author} {\bibfnamefont {P.}~\bibnamefont {Chaudhuri}}, \
  and\ \bibinfo {author} {\bibfnamefont {S.}~\bibnamefont {Vemparala}},\
  }\bibfield  {title} {\enquote {\bibinfo {title} {Effect of ring stiffness and
  ambient pressure on the dynamical slowdown in ring polymers},}\ }\href@noop
  {} {\bibfield  {journal} {\bibinfo  {journal} {Soft Matter}\ }\textbf
  {\bibinfo {volume} {18}},\ \bibinfo {pages} {2959--2967} (\bibinfo {year}
  {2022})}\BibitemShut {NoStop}%
\bibitem [{Note1()}]{Note1}%
  \BibitemOpen
  \bibinfo {note} {We note that while average asphericity increases with the
  ring stiffness, there is a reversal at higher values of $K_{\theta }$ (data
  not shown) when the rings become circular in shape and the asphericity then
  nearly vanishes.}\BibitemShut {Stop}%
\bibitem [{\citenamefont {Berthier}(2011)}]{berthier2011dynamic}%
  \BibitemOpen
  \bibfield  {author} {\bibinfo {author} {\bibfnamefont {L.}~\bibnamefont
  {Berthier}},\ }\bibfield  {title} {\enquote {\bibinfo {title} {Dynamic
  heterogeneity in amorphous materials},}\ }\href@noop {} {\bibfield  {journal}
  {\bibinfo  {journal} {arXiv preprint arXiv:1106.1739}\ } (\bibinfo {year}
  {2011})}\BibitemShut {NoStop}%
\bibitem [{\citenamefont {Kawasaki}, \citenamefont {Araki},\ and\ \citenamefont
  {Tanaka}(2007)}]{kawasaki2007correlation}%
  \BibitemOpen
  \bibfield  {author} {\bibinfo {author} {\bibfnamefont {T.}~\bibnamefont
  {Kawasaki}}, \bibinfo {author} {\bibfnamefont {T.}~\bibnamefont {Araki}}, \
  and\ \bibinfo {author} {\bibfnamefont {H.}~\bibnamefont {Tanaka}},\
  }\bibfield  {title} {\enquote {\bibinfo {title} {Correlation between dynamic
  heterogeneity and medium-range order in two-dimensional glass-forming
  liquids},}\ }\href@noop {} {\bibfield  {journal} {\bibinfo  {journal}
  {Physical review letters}\ }\textbf {\bibinfo {volume} {99}},\ \bibinfo
  {pages} {215701} (\bibinfo {year} {2007})}\BibitemShut {NoStop}%
\end{thebibliography}%
\end{document}